# Atomic-scale investigation of the irradiation-resistant effect of symmetric tilt grain boundaries of Fe-Ni-Cr alloy

Zhijia Liu[a], Zehua Feng[a], Heran Wang[b], Yunpeng Zhang[a], Zheng Chen[b], Fanchao Meng[c, *], Jing Zhang[b, *]

[a] *School of Material Science and Engineering, Xi'an University of Technology, Xi'an, Shaanxi 710048, China*

[b] *School of Material Science and Engineering, Northwestern Polytechnical University, Xi'an, Shaanxi 710072, China*

[c] *Institute for Advanced Studies in Precision Materials, Yantai University, Yantai, Shandong 264005, China*

**Abstract:** The development of efficient next-generation nuclear reactors is necessary to relieve the contradiction of rising energy consumption and fossil energy depletion. The operation security of reactors depends on the microstructure and mechanical properties of reactor core materials upon harsh radiant, high-temperature, and high-pressure conditions. In this paper, the Fe-20Ni-25Cr alloy that is used for fuel cladding or pressure vessels with various grain boundaries (GBs) was investigated by employing molecular dynamics simulations. The bi-crystals comprised of $\Sigma3(111)$, $\Sigma3(112)$, $\Sigma9(114)$, $\Sigma11(113)$, $\Sigma19(116)$, and $\Sigma17(223)$ types GBs were considered to systematically examine the interplay between irradiation defects, irradiation microstructure evolution under stress, and irradiation mechanical properties with irradiation intensity, coincidence site lattice parameter, tilt angle, and GB thickness. It is found that irradiated vacancies and interstitials are annihilated by competitive GB absorption and recombination. Bias absorption of interstitials is observed for most bi-crystals except $\Sigma3(111)$ and $\Sigma11(113)$ at 15 keV incident energy, and results in abundant residual vacancies clusters in grain interior. In addition, different GBs exhibit quite diverse irradiation defect sink ability, and the number of residual vacancies is inversely related to the GB thickness, where $\Sigma3(111)$ and $\Sigma11(113)$ GBs with narrow GB thickness are weak in defect absorption and the others are strong. Furthermore, uniaxial tensile simulations perpendicular to the GB reveal that all of the mechanical performance of bi-crystals deteriorates after irradiation, which originates from dislocation propagation facilitated by irradiation defect clusters. In particular, regardless of whether the irradiation is applied, the maximum tensile strain, toughness, and Young's modulus are monotonically correlated with GB's tilt angle, while the ultimate tensile strength is stable for larger GB's CSL parameter. Finally, on the basis of the evolution of the irradiation defects, microstructures, and mechanical performances, we proposed guidelines of rational design of irradiation-resistant Fe-Ni-Cr alloy.

* Corresponding authors.
E-mail address: jingzhang@nwpu.edu.cn (J. Zhang); mengfanchao@ytu.edu.cn (F.C. Meng).

# 1 Introduction

During the operation of nuclear reactors, the core materials of reactors suffer from persistent irradiation, harsh temperature, and rigorous pressure. The FCC concentrated solid-solution Fe-Ni-Cr alloys are the critical materials that serve as fuel cladding or pressure vessel materials in nuclear reactors due to their excellent thermal conductivity, mechanical properties, and radiation resistance [1,2]. With the increasing demand for nuclear safety and economical efficiency, the next generation reactors are expected to operate at a higher temperature, higher pressure, and even severe energetic irradiation, which challenges the present alloys used. A comprehensive understanding of the relationship among irradiation dose, defects generation, defects annihilation, and intrinsic GB sinks will benefit the materials design with enhanced balance properties. The incident energetic particles cause cascade displacement, generating considerable vacancies and interstitials in the crystal lattice, which would segregate on GBs or other sinks, or would evolve into larger size defects cluster, such as dislocation loops, voids, and stacking fault tetrahedrons (SFT)[3,4,5]. The formation of these defective microstructures is believed the main reason cause structural instability, and thereafter, leads to stress concentration on the surface of GBs or other defective microstructures. In considering the thermal effect and mechanical effect, the materials may present irradiation softening, hardening, and embrittlement, which leads to material failure in severe cases[6].

The defective microstructures depend on the irradiated vacancies and interstitials level, therefore, the reduction or control of these defects has an essential significance in the rational design of radiation tolerance materials. These irradiated point defects could be eliminated by vacancy-interstitial recombination or intrinsic sinks like dislocations, precipitates, or grain boundaries (GBs)[7]. GBs are common microstructural defects in materials, and the vacancies and interstitials generated by the radiation tend to diffuse to the GBs, and thus GBs are considered to be key radiation defect sinks[8-9]. J. Han[10] pointed out that the ability of GBs to catch defects varies depending on the GB structure. B.N. Singh[11] recognized that because of the defects captured by the GBs, the radiation damage decreases with the grain size reduction. Therefore, the effect of irradiation on the mechanical properties of polycrystalline materials can be reduced by reasonably designing the GB ratio and microstructure; eventually, the degree of radiation hardening and embrittlement can be reduced. For example, nano metals, nanoporous materials, and nano multi-layer materials with a high GB ratio exhibit excellent radiation damage resistance than traditional metals[12].

To reveal the mechanism of interaction between GBs and point defects, scholars conducted a large number of cascade displacement simulations. They found that the GB exerts a bias-absorption effect on the interstitials compared with vacancies. The formation energy of vacancies and interstitials at the GB is lower than that in the grain interior (GI), and the formation energy of interstitials at the GB is lower than that of vacancies[13]. X. Zhang et al.[14] indicated that the vacancy migration energy in FCC metals varies from 0.7 to 1.7 eV, and the interstitials migration energy is much smaller than that of the vacancy, which is typically 0.05-0.1 eV. Interstitials at room temperature show comparatively high mobility, whereas the vacancies mostly move at elevated temperatures. Because of the low formation energy and high mobility of interstitials, plentiful interstitials are accumulated in the GB area, and most vacancies are left in the grain interior (GI) after irradiation. Subsequently, the researchers also found that the interstitial-rich GB will reduce the energy barrier for the migration of interstitials so that the GB emits interstitials to eliminate the vacancies near the GB[15,16]. Although the GB can release interstitials to recombine with vacancies in the GI, it will not change the GB region's interstitial enrichment and the vacancy enrichment in the GI. The efficiency of GB absorption of point defects is related to many factors. The further the distance between the primary knock-on atom (PKA) and the GB, the more the number of surviving vacancies[17]. In the body-centered cubic (BCC) metal, D. Chen[18] pointed out that the high-angle GB can provide more low defect formation energy sites and more "grain boundary chain" defect formation direction than the low-angle, which is more conducive to the annihilation of defects.

After capturing the point defect, the microstructures change, and the corresponding mechanical properties are different from the original GB; the captured interstitials or vacancies can reduce the average sliding friction of symmetric tilt GB under shear deformation by about an order of magnitude, and GB migration may occur[19]. Of course, each GB has its unique structural characteristics, surface energy and diffusion coefficient, and its ability to capture point defects is also different. After irradiation, the number and aggregation state of point defects in the crystal will also be diverse, and its macroscopic mechanical properties will be different[20,21]. A. Kedharnath[22] compared the number of point defect and mechanical properties of GBs after irradiation of $\Sigma3$, $\Sigma11$, low angle GB (LAGB), and high angle GB (HAGB) in α-Fe, and obtained the order of the UTS following: $\Sigma3$ > LAGB > HAGB > $\Sigma11$. The UTS loss of $\Sigma3$ GB is severe because it cannot act as defect sinks under irradiation, and the LAGB grain boundary is an effective defect sink. X.Y. Wang et al.[23] studied four kinds of symmetric tilt GBs of $\Sigma3(111)$, $\Sigma3(112)$, $\Sigma5(012)$ and $\Sigma11(332)$ in BCC iron and found that the UTS decreased after irradiation owing to the rearrangement of atoms, the formation of stress concentration and vacancy enrichment zones, and the nucleation and movement of dislocation under stress. These results indicate that the macroscopic properties have a specific relationship to absorb point defects and GB structure. To further explore the radiation-resistant potential of Fe-Ni-Cr alloys, it is necessary to understand the defects' sink efficiency of various GBs and the corresponding mechanical properties.

So far, although much basic research has been carried out to explore the effect of grain boundaries on the mechanical behavior of metals, there are relatively few reports on the mechanical properties of irradiated GB in FCC alloys, especially for single-phase concentrated solid-solution alloy. GB Engineering proposes that FCC metals with low and medium stacking fault energy (such as nickel-based alloy and austenitic stainless steel) can increase the proportion of special GBs through deformation and heat treatment, which can significantly improve the macroscopic properties of materials. Among them, the most commonly used processing method for FCC metals is to add a large number of annealing twins to the alloy, that is, $\Sigma3$ GBs, in which the migration of non-coherent $\Sigma3$ GBs will generate a large number of $\Sigma9$ GBs, the content of which is up to 10%. Many studies have been reported for irradiation damage in the $\Sigma3$ symmetric tilt GBs. Therefore, this article conducts a more detailed study on the $\Sigma9$ GB in the FCC single-phase concentrated solid-solution alloy.

Molecular dynamics (MD) simulation is a powerful tool to study the effect of GB structure and

GB types on irradiation and mechanical properties of materials at the nanometer scale[24]. In this paper, MD simulation was adopted to study the cascade displacement and tensile properties of the FCC single-phase concentrated solid-solution alloy of single crystal (SC) and Σ9(114) bi-crystal attacked by primary knock-on atom, to discuss the similarities and differences of irradiation damage between SC and bi-crystal. Then, systematic simulations of irradiation damage and tensile properties of SC and six symmetric tilt GBs, being Σ19(116), Σ17(223), Σ11(113), Σ9(114), Σ3(112), and Σ3(111) were performed to reveal the relationship between the GB characteristics, such as the coincidence site lattice (CSL) parameter (Σ), tilt angle, and GB thickness, and the material's radiation damage resistance. This study can help clarify the effects of GBs on the radiation damage resistance of the FCC single-phase concentrated solid-solution alloy, offering important insights into the radiation damage research of multicomponent alloys, especially high/medium entropy alloys.

## 2 Calculation and research methods

### 2.1 Simulation methodology

The cascade displacement of Fe-20Ni-25Cr single-phase concentrated solid-solution alloy was simulated using the MD method, carried out by Large-scale Atomic Molecular Massively Parallel Simulator (LAMMPS). This article uses the embedded atom method (EAM) potential developed by Bonny et al.[25] in 2013. This potential is designed to model radiation-induced defects in FCC FeNiCr. It could properly predict the atom's interaction under radiation damage[26]. In the cascade simulation, the short-range repulsive interaction was described by Ziegler-Biersack-Littmark (ZBL) potential.

The calculation model is FCC structure, with alloy composition being 55%Fe, 20%Ni, and 25%Cr, about 15nm in three dimensions, and 270,000 total atoms. The effect of irradiation is tested on a single crystal (SC) and Σ9(114) bi-crystal model, with the primary knock-on atom (PKA) having incident energy ($E_{pka}$) of 2, 5, 10, and 15 keV. The mechanical properties after irradiation of SC and Σ9(114) bi-crystal were compared after the uniaxial tension test. A general description of irradiation microstructures links to the various symmetrical tilt GBs at 15KeV incident energy are given, these GBs are Σ3(111), Σ3(112), Σ9(114), Σ11(113), Σ17(223), and Σ19(116). To ensure the accuracy of the simulation, each sample that suffers a certain $E_{pka}$ was simulated five times.

### 2.2 Grain boundary models and descriptions

In this paper, GB models were constructed based on the CSL model. The GB structure is formed by two grains rotating at a specific angle in the opposite direction with [110] as the rotation axis. For example, in the Σ9[110](114) GB model, the two grains use [110] as the tilt axis, the upper and lower grains rotate 19.47°clockwise and counterclockwise to form a new relationship. Among them, (114) plane refers to the common interface created by the rotation of the two grains: the GB plane; Σ9 represents that the coincident site density is 1/9. For each GB (such as Σ9(114)) type, atoms at the grain boundary areas can be arranged in various ways (7 kinds)[27,28]. Fig.1. shows the schematic diagram of all the 6 kinds of GB models, that is, Σ3(111), Σ11(113), Σ9(114), Σ17(223), and Σ19(116)，and Σ3(112), Conjugate gradient (CG) algorithm was employed in the

MD simulation for energy minimization of each structure, and then the structure with the lowest GB energy was used as the target GB model.

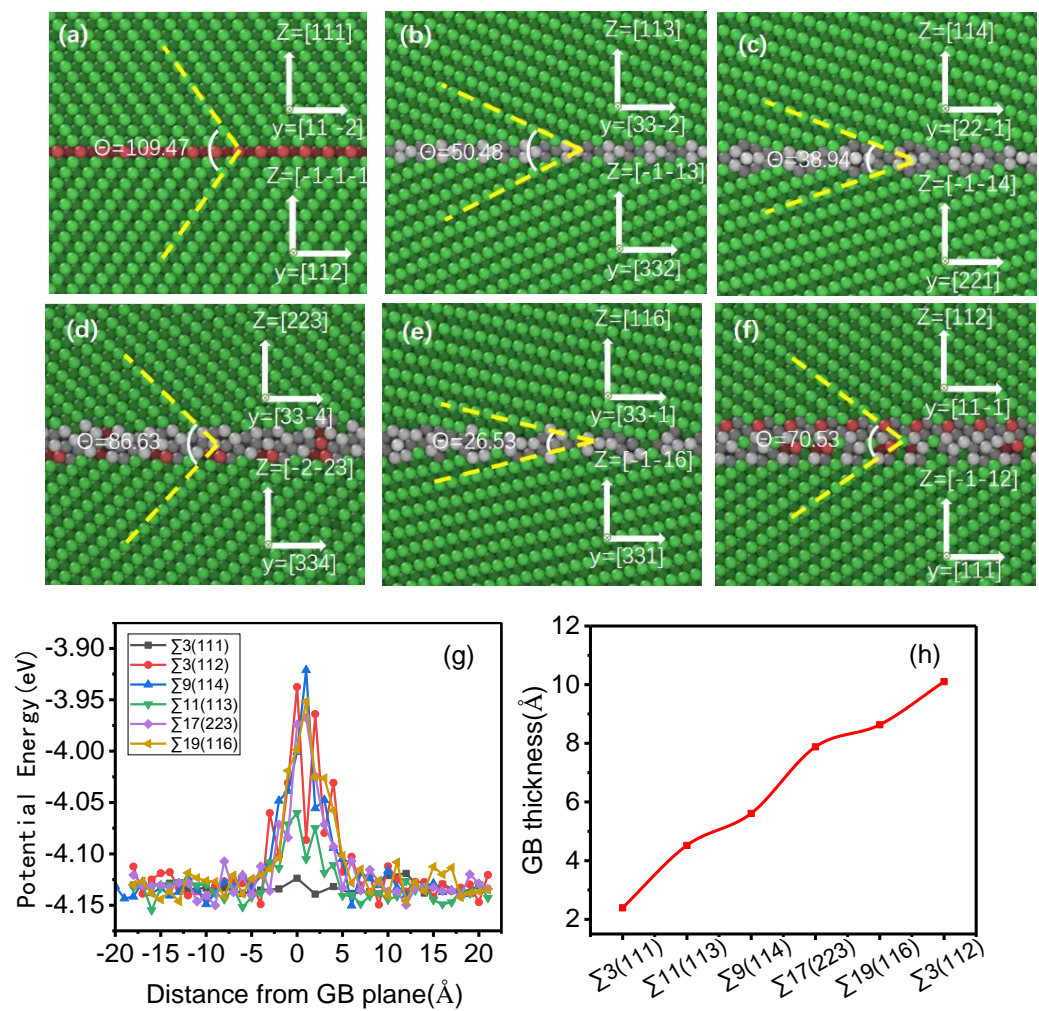

Fig. 1 Bi-crystals structures of different GBs in (a) Σ3(111), (b) Σ11(113), (c) Σ9(114), (d) Σ17(223)), (e) Σ19(116), and (f) Σ3(112). (g) The average potential energy of various GBs, and (h) GB thickness-GB types relation. The GB atoms are colored gray, HCP atoms are colored red, and atoms inside the grain (i.e., FCC atoms) are colored green.

Because the energy of atoms in the GB area is different from those in the GI. The average potential energy($E_{ave}$) of atoms at different distances from the GB is calculated by $E_{ave}=E_{layer}/N_{layer}$, where $E_{layer}$ is the total energy and $N_{layer}$ is the total atom number of the layer. The average potential energy across the double sides of the GB is given by Fig. 1(g), in which the "0" point denotes the center of a GB. The atomic potential energy that locates in the middle of GB is higher than that deviates from the GB, thus, it fluctuates down from the middle, converges around -4.12 eV on both sides of GB. The fluctuation contributes to the inhomogeneous distribution of Fe, Ni, and Cr elements on each slab. Particular attention should be paid to Σ3(111) and Σ11(113), they have narrow GB width, and not much discrepancy of atomic potential energy on GB and grain interior(GI). The above results are in line with the results reported by Wang et. Al.[23] Each GB can be distinguished by tilt angle and GB thickness, the tilt angle is demonstrated as Fig. 1(a-f), and Fig. 1(h) is the GB thickness measured across the grey region. GBs are effective sinks for irradiation point defects to maintain microstructure stability, we will check how they work in the following.

**2.3 Irradiation and uniaxial tensile simulation**

Periodic boundary conditions (PBC) were used for simulations of cascade displacement evolution. The systems were first relaxed using the CG method to find the local minimal energy state. After achieving the minimum energy configuration, equilibration at the desired temperature (300 K) was performed for 20 picoseconds (ps) with a time step of 1 femtosecond (fs). During the equilibration phase, the simulation box was subjected to an NPT ensemble. In the cascade simulation, as shown in Fig. 2, a random iron (Fe) atom at approximately 37 Å below the GB was selected as the PKA atom, which has a certain initial kinetic energy with its velocity perpendicular to the GB plane. The time-step is automatically adjusted from $10^{-3}$ fs to 1 fs with the displacement of each atom within each time step not exceeding 0.005 Å during irradiation damage[29]. In the cascade simulation, the NVT and NVE ensemble is used for the outermost three lattice and inner atoms[30]. For statistical accuracy, each GB system is simulated at least five times with various random number seeds for atomic velocity initialization in MD simulations.

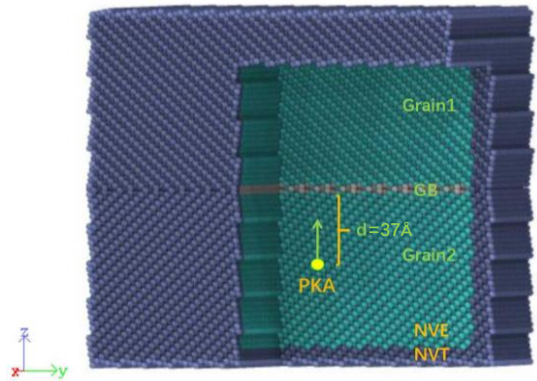

Fig. 2 Schematic diagram of irradiation simulation. The gray area in the middle is the GB, yellow is the PKA atom, and d is the distance from PKA to the GB plane. The outer three lattice layers were equilibrated using the Nose-Hoover thermostat (NVT ensemble) at 300 K throughout the cascade displacement simulation to act as a heat sink. The internal atoms use the NVE ensemble for dynamic evolution.

The tensile properties of different GBs under zero pressure at room temperature (300 K) were calculated. Before the tensile simulation, the irradiated model must be fully relaxed with the NPT ensemble. The tensile force perpendicular to the GB plane, a strain rate of 0.0005/ps, and a time step of 1 fs were used. PBC was kept for all three dimensions and the stress in the x and y directions remains at 0 Pa during the simulation.

The Open Visualization Tool (OVITO) software was used for post-processing[31]. The Wigner-Seitz method was applied to analyze the type and number of point defects. The Common Neighbor Analysis method was employed to investigate the crystal structural transformation in the cascade. The Dislocation Extraction Algorithm (DXA) was employed to analyze the properties of dislocations[32,33].

**3 Results and discussion**

Molecular dynamic simulation was performed on Σ3(111), Σ3(112), Σ9(114), Σ11(113), Σ19(116), and Σ17(223) bi-crystals, and an SC sample is also included as a reference. The Σ9(114) sample with medium tilt angle is selected to study the effects of GB when exposed to 2, 5, 10, and 15 keV irradiation conditions. Comparison of SC, Σ3(111), Σ3(112), Σ9(114), Σ11(113), Σ19(116), and Σ17(223) bi-crystals from various aspects are given to clarify how each GB interacts with irradiation, also, the contribution from coincidence site lattice parameter, tilt angle, and grain boundary thickness are analyzed. Finally, a design strategy is proposed.

**3.1 Irradiation defects generation and annihilation of Σ9(114)**

Vacancies and interstitials are created by cascade displacement caused by incident energetic particles and accumulate with persistent exposure to the irradiation condition, they are the origins of disastrous damages like voids, dislocation loops, or bubbles. Thus, the statistical peak number of point defects of Fe-Ni-Cr alloy that includes a Σ9(114) type bi-crystal under 2, 5, 10, and 15keV PKA energy are listed in Table1, the results for SC are also included as a reference. The peak number of point defects is dependent on the PKA energy, it is getting concentrated with enhanced PKA energy for both SC and Σ9(114). During such cascade peak stage, there is not much difference between SC and Σ9(114) of their peak defects number. We further examine the evolutional point defects of SC and Σ9(114) bi-crystal showed in Fig. 3.

Table 1 The peak number of interstitials, which represents the maximum number of defects (interstitial or vacancy) in the cascade displacement process. Note that the peak number of interstitials equal to that of vacancies. The simulation time should be at least 50ps.

|         | 2keV       | 5 keV        | 10 keV         | 15 keV         |
|---------|------------|--------------|----------------|----------------|
| SC      | 42±7.48    | 139±26.39    | 1565±304.06    | 3528±578.97    |
| Σ9(114) | 44±16.27   | 357±117.23   | 1978.±378.16   | 3556±495.93    |

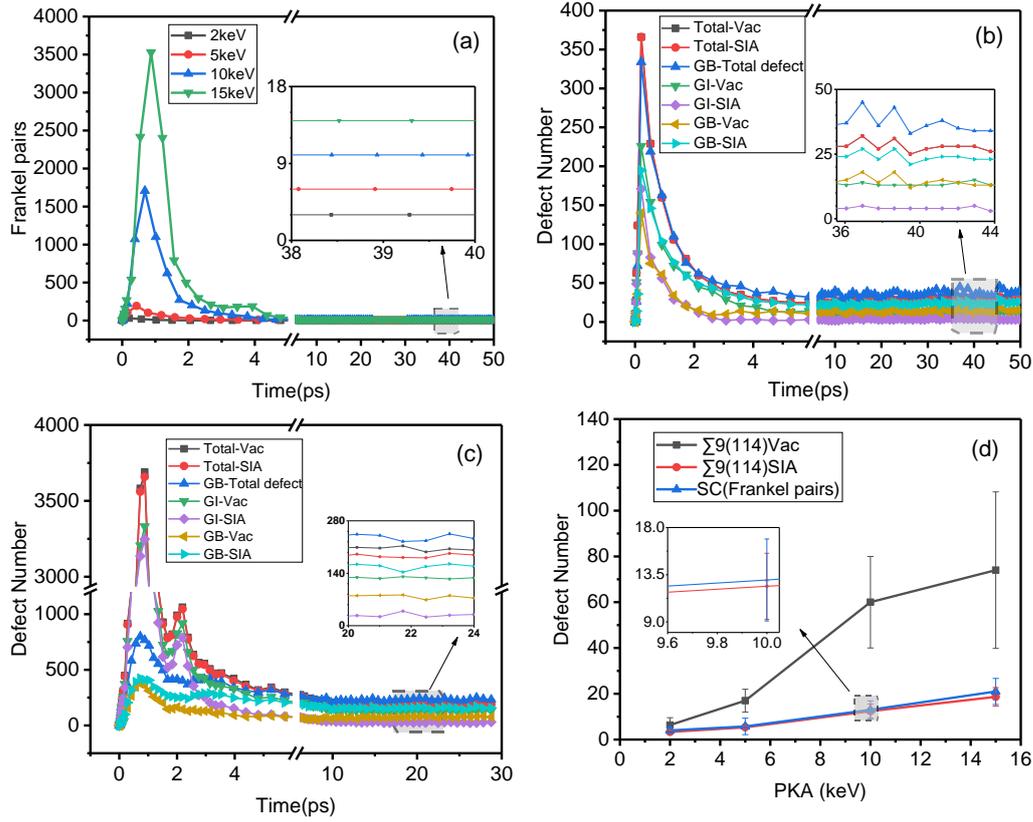

Fig. 3 The number of self-interstitial atoms and vacancies evolution with the simulation time for SC and Σ9(114) bi-crystal: (a) SC defects number (interstitials or vacancies) evolve with time. (b) Defect number evolution of Σ9(114) bi-crystal with PKA energy equal to 5 keV. (c) Defect number evolution of Σ9(114) bi-crystal with PKA energy equal to 15 keV. (d) Comparison of the number of defects at different PKA energies between SC and Σ9(114) bi-crystal in the GI.

The primary damage events pathway caused by incident energetic particles, including PKAs, high order knock-on atoms, cascade displacement, thermal spike, and quenching. Defects evolution of Fig.3a demonstrates this process. During the cascade displacement, PKA transfers energy to the surrounding lattice atoms, and the number of defects surging to the maximum. Meanwhile, the temperature rises rapidly in the cascade region due to the high velocity of interstitials. Subsequently, the heat in the cascade area is transferred to the surrounding lattice atoms, so the temperature starts to decrease, the vacancies and interstitials recombine, therefore the number of Frankel pairs gradually decreases, as shown in Fig.3a for SC, Fig. 3b and 3c for Σ9(114) bi-crystal. Finally, the number of defects in the crystal tends to be stable. As there are no defect sinks of SC, vacancies and interstitials always appear in pairs (the number of vacancies is equal to the number of interstitials). We compare the average number of surviving Frenkel pairs (FPs) in SC with the previous reports obtained at 10 keV, we have 12 surviving Frenkel pairs in our work, for Ni-50Fe[34], pure Ni[35], Ni-based alloys[36] the number is 10, 18, and 9 respectively. Our results are in general agree with the former work.

The stable defects number in the SC and Σ9(114) bi-crystal as a function of PKA energy is plotted using the blue line in Fig. 3d. The defect number is closely related to PKA energy, it concentrates as the enhanced PKA energy. The differences lie in that the defects are Frenkel pairs in SC, and inequitable vacancies and interstitials in Σ9(114) bi-crystal. The vacancy concentration is times of interstitial concentration in Σ9(114) bi-crystal, the interstitials are mobile than vacancies, Σ9(114) type grain boundary is effective sinks of interstitials. On the other hand, the faster mobility of interstitials toward Σ9(114) type GB sinks leaves abundant vacancies in the GI, thus the probability of recombination of vacancies and interstitials lessens. Vacancies and interstitials recombine in the SC, the residual vacancies and interstitials appear in pairs and hold a comparatively dilute level. Vacancies-interstitials recombination and vacancies-interstitials GB-adsorption occur simultaneously in the Σ9(114) bi-crystal, the residual vacancies are much heavier than interstitials due to GB's bias adsorption. These residual interstitials in Σ9(114) bi-crystal and SC hold a similar level of all the irradiation condition investigated, the Σ9(114) type GBs is an effective and strong interstitial sink, it could accommodate as many interstitials as the PKA energy reach 15 keV.

GBs are intrinsic sinks for irradiation defects, they absorb vacancies and interstitials, which could annihilate defects efficiently and thus lessen the possibility of forming abundant defect clusters, therefore, the existence of GBs stabilizes the microstructures on one hand. On the other hand, bias absorption reduces the possibility of vacancies-interstitials recombination leaving abundant residual vacancies and bits of interstitials. The residual interstitials are minor, then it may be accommodated by the crystal lattice without cause severe lattice distortion or may segregate on the close stacking plane forming interstitial loops[37,38,39]. The residual vacancies are major defects, they diffuse sluggishly, rearrange along the close stacking direction forming dislocations or stacking fault tetrahedron, or they may nucleate and grow into voids, all of which have been observed by the experiments[40,41,42]. The mechanical stress will accelerate the stated process, and we will investigate how tensile stress interacts with these defects in the following.

**3.2 Microstructure evolution of Σ9(114) bi-crystal under tensile test**

Fig. 4 shows the stress-strain relation and dislocation density of both SC and Σ9(114) bi-crystal under different PKA energies. For stress-strain statistics, the maximum point of the stress-strain curve is defined as the ultimate tensile strength (UTS), and the strain corresponding to the UTS is the maximum tensile strain (MTS). The UTS and MTS under different PKA energies are concluded in Fig. 4e. Young's modulus is obtained by fitting the stress-strain curve using the Hooke's law $\sigma=E*\varepsilon$ in the linear correlated stage within 0.025 strain, where $\sigma$ is stress, $\varepsilon$ is the strain, and E is Young's modulus of the material. The Young's modulus at different PKA energy is shown in Fig. 4f.

The stress-strain response of SC is shown in Fig. 4a, where we see that $\sigma$ and $\varepsilon$ are linearly correlated exhibiting elastic characteristics for a larger strain range. In considering the irradiation effects, the UTS declines steadily for each enhanced PKA energy, because of the linear relation of UTS and MTS, the MTS also shifts left gradually. The UTS of unirradiated SC is 11.79 GPa, and the UTS is about 10.56 GPa when PKA is 15 keV, which is about 10.43% lower than that of unirradiated material. The stress-strain curve of Σ9(114) bi-crystal is shown in Fig. 4c, the introduction of Σ9(114) GB in the material decreases the UTS comparing with SC, but its UTS only gets reduced at severe irradiation condition of 15 keV (UTS$_{15keV}$ = 7.01 GPa), with a 17.24% decrease in comparing with that at 0 keV (UTS$_{0keV}$ = 8.47 GPa).

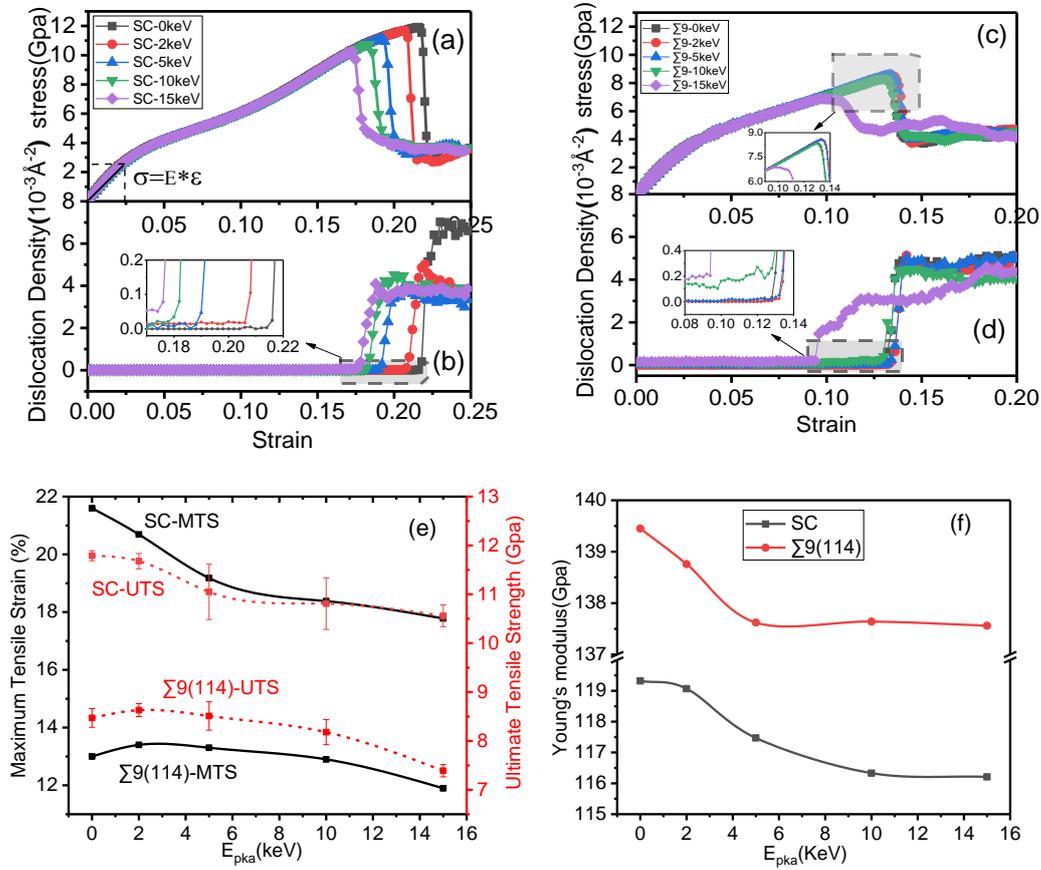

Fig. 4 Tensile stress-strain curves and dislocation density-strain relation for (a-b) SC and (c-d) Σ9(114) bi-crystal under different PKA energies, respectively. The mechanical properties of SC and Σ9(114) bi-crystal, (a) UTS-E$_{pka}$ relation and MTS-E$_{pka}$ relation, and (b) Young's modulus.

To explore the relationship between PKA energy and UTS, the dislocation line density was adopted. At small PKA energy, only a small amount of point defects or clusters are present in the crystal, and these clusters preferentially evolve into dislocations during the tensile process. Fig. 4b and 4d show the evolution of dislocation density under different PKA energies for SC and Σ9(114) bi-crystal. During the tensile process, the dislocation density remains unchanged until the UTS is reached, after that, dislocations are produced and get the maximum in a short time. The maximum dislocation density and the UTS almost appear simultaneously, with the occurrence time of UTS is slightly earlier than that of maximum dislocation density. The propagation of dislocations soon after the UTS makes the materials fail. It should point out that, case of the Σ9(114) bi-crystal under 15 keV incident energy is different, besides that the UTS and MTS are drastically lower than that at the 10 keV sample, dislocations propagate before the materials attain its UTS, which is believed to be related to the high vacancies level in GI and interstitials aggregation in GB. Besides, by calculating dislocation densities under different PKA energy, it was found that the dislocation density in this work was about $10^{-20}$-$10^{-19}$ cm$^{-2}$, far lower than the dislocation density in the actual crystal ($10^6$-$10^{12}$ cm$^{-2}$). In such a minor dislocation density range, the crystal strength decreases with the increase of dislocation density as it is believed that atoms are movable in a crystal containing dislocations than in a perfect one. Dislocations facilitate deformation when they are not severely interacting with each other.

As observed in Fig.4e, the SC sample exhibits superior strength and ductility than the Σ9(114) bi-crystal sample, and the Σ9(114) GB itself causes reduction of the strength and ductility compared with SC. Although there are abundant vacancies left in the GI of Σ9(114), it seems that the UTS and MTS are insensitive to incident energy within 10 keV. The Young's modulus of SC and Σ9(114) bi-crystal decreases slightly with the enhanced PKA energies, the young's modulus of Σ9(114) bi-crystal is higher than that of SC.

Comparing the dislocation density of Σ9(114) with SC, the average dislocation is denser, the UTS is inferior. The enhanced irradiation accelerates dislocation propagation in SC. For Σ9(114) bi-crystal, the dislocation under 15 keV PKA energy grows in two stages: The first starts from around 0.09 strain and the second starts from about 0.12 strain where the UTS reached. Σ9(114) GB absorbs most interstitials leaving abundant vacancies in GI (Fig. 3d), these vacancies rearrange along with a certain special direction cause dislocations propagation in the first stage. Then, we investigated GB's effects on the defective microstructure evolution of irradiated samples with stress perpendicular to GB; the dislocation evolution of SC and Σ9(114) bi-crystal after 10 keV irradiation was shown in Fig. 5.

The irradiated point defects are generated in the SC, which is dispersed or aggregated into clusters. As there is a constant strain rate applied, the defect clusters evolve into a 1/6<112> Shockley dislocation until the strain reaches 0.174 (Fig. 5a), or, evolve into 1/2<110> perfect dislocation (Fig. 5b). Dislocations are rare till the strain attains 0.186, where the fracture occurs; generating a large number of 1/6<112> Shockley dislocations around it (Fig. 5c). The load was keeping applied till strain up to 0.192, and the dislocation density grows abruptly (Fig. 5d). Dislocations nucleate on the defect clusters, because of the limited irradiated defects in SC, dislocation propagation is relatively slow and always persists at a low-density till fracture.

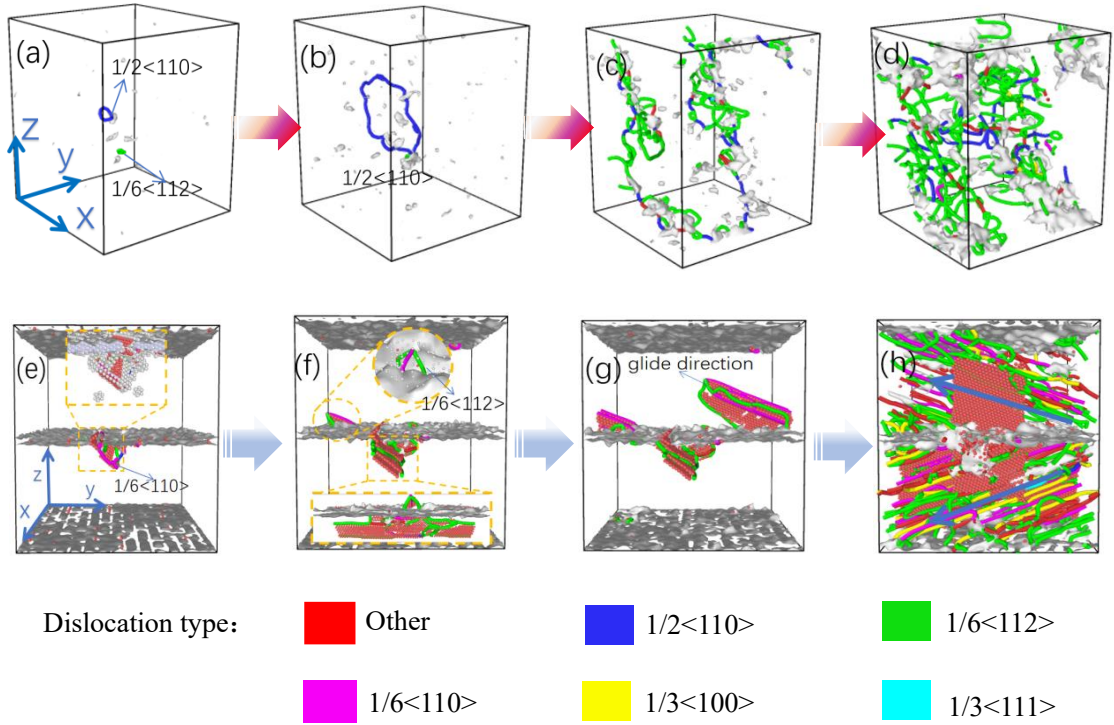

Fig. 5 Dislocation evolution of a SC (a-d) and Σ9 bi-crystal (e-f) irradiated at 10 keV under different strains: (a) 0.174, (b) 0.184, (c) 0.186, (d) 0.192, (e) 0, (f) 0.122, (g) 0.132, and (h) 0.134. The FCC atoms were hided. The HCP atoms color red. The colored lines represent different dislocation lines.

For the Σ9(114) bi-crystal, the defects produced by irradiation are unlikely to traverse the GB for 10 keV PKA energy. It is found that Most interstitials diffuse into GB, most vacancies distribute on one side of GB and evolve into vacancy clusters, and a stacking fault tetrahedron (SFT) evolve from the aggregation of vacancies (Fig. 5e) (note, no stacking fault tetrahedron was observed in the SC under the same PKA energy). Tensile stress accelerates the rearrangement of vacancies parallel to GB and then propagates dislocations, where these dislocations extend directionally, and the 1/6<112> Shockley and 1/6<110> Stair-rod dislocations are observed, as shown in Fig. 5f. As the tensile stress gradually increases, GB emits dislocations to the grain on the other side, and the whole sample is filed with dislocations momentarily, see Fig. 5g-h.

From the tensile dislocation's evolution diagram of SC and Σ9 bi-crystal, it can be seen that the deformation mechanism of the two is different. Point defects in the SC are uniformly distributed in the center of the model in the form of vacancies or interstitials. In the Σ9 bi-crystal, the defects are mainly the vacancy clusters in the GI and the interstitials in the GB region. The vacancy clusters in the GI are usually larger and more densely distributed than those in the SC and even evolve into a stacking fault tetrahedron, making it easy to cause stress concentration and acting as the first source of dislocation. Under tensile stress, dislocation nucleation occurs earlier at GB than that in SC, which reduces UTS. The evolution of dislocations in the SC is dominated by 1/2<110> perfect dislocation slipping in the {110} crystal plane; while dislocation slip is divided into two parts in Σ9 bi-crystal, the first part is 1/6<112> Shockley forming by vacancy clusters inside the grain and slipping in {111} crystal plane along the <110> crystal direction, the second part is 1/6<110> Stair-rod dislocations nucleation at the GB and evolution in {001} crystal plane. In short, due to the existence of GBs, large vacancy clusters in the GI cause stress concentration, and dislocations nucleate at the GB in advance during the tensile process, which reduces the UTS, so the UTS of Σ9 bi-crystal is lower than that of SC. The effects of other kinds of GB are further investigated.

**3.3 General law of irradiation microstructures and properties of various GBs**

Fig. 6 provides the number of defects of different GB types after 15 keV irradiation. Various GBs show significant difference against irradiation, materials with Σ3(111) and Σ11(113)GBs, whose irradiation defects annihilate mostly by recombination of vacancies and interstitials, are almost free of defects compared with the other ones. Σ3(111) and Σ11(113) GBs are weak defects sinks, while Σ3(112), Σ9(114), Σ17(223), and Σ19(116) are strong defects sinks. The total defects increase in order as follows, SC, Σ3(111), Σ11(113), Σ9(114), Σ17(223), Σ19(116), Σ3(112). The interstitials in GB (GB-SIA, the green line), the interstitials in GI (GI-SIA, the yellow line), the vacancies in GB (GB-Vac, the blue line), the vacancies in GI (GI-Vac, the purple line), these four all hold the same order. The interstitials numbers in all kinds of bi-crystal samples in Fig.6 are comparatively small.

By contrasting Fig.1h and Fig. 6, we find that the defect number monotonically increases with the GB thickness widening. The defect recombination and GB absorption are competitive; bias adsorption of interstitials reduces recombination probability, rises residual vacancies in GI; A thicker GB, such as Σ9(114), Σ17(223), Σ19(116), and Σ3(112), are efficient in bias adsorption. A weak GB sink, whose irradiated defects were annihilated mostly by recombination, is not efficient enough. A strong GB sink could annihilate irradiated defects by both recombination and GB absorption, then, it is effective to stabilize the microstructures on one hand; on other hand, the residual vacancies in GI will nucleate and evolve into voids or dislocations, accelerate failure. The balance of efficiency, microstructures, and mechanical properties are complicated, a rational design of Fe-Ni-Cr alloy should consider both GB types and GB areas to get durable irradiation stability.

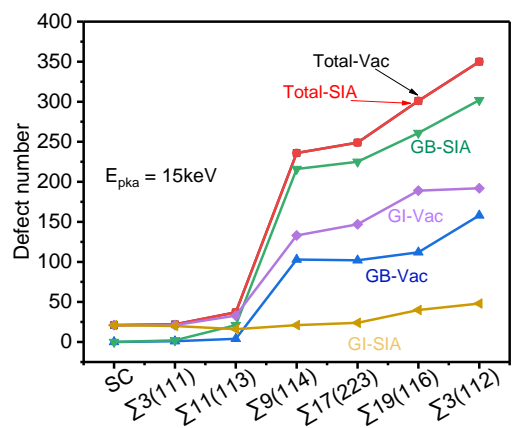

Fig. 6 The statistics irradiation defects number of different GB at 15 keV (Note: Total-Vac: the total number of vacancy; Total-SIA: the total number of interstitials; GB-Vac: vacancy atoms number in the GB; GI-Vac: vacancy

atoms number in the GI; GB-SIA: interstitials number in the GB; GI-SIA: interstitials number in the GI).

The effects of irradiated defects are further explored through mechanical performance and dislocation evolution. Fig. 7 (a) and (b) shows the stress-strain relation of various GBs and SC at 15keV PKA energy. SC is ductile and strengthful, bi-crystal containing GB loses strength and ductility which the specifics depend on the GBs' microstructure. Among them, the Σ3(111) bi-crystal has the maximal UTS and minimal MTS. Comparing Fig.7a with Fig.7b, the UTS reduces and MTS shifts left for each GB after irradiation; that is, irradiation softening occurs, which is in agreement with the tensile results on symmetric tilt GB by A. Kedharnath[43] and X.Y. Wangs[44]. X.Z. Xiao[45] pointed out that there is a critical grain size of 27nm for radiation softening undersize or hardening oversize. The size of this article is about 15 nm in three dimensions, which is much less than 27 nm, that is why only irradiation softening was observed for the present simulation.

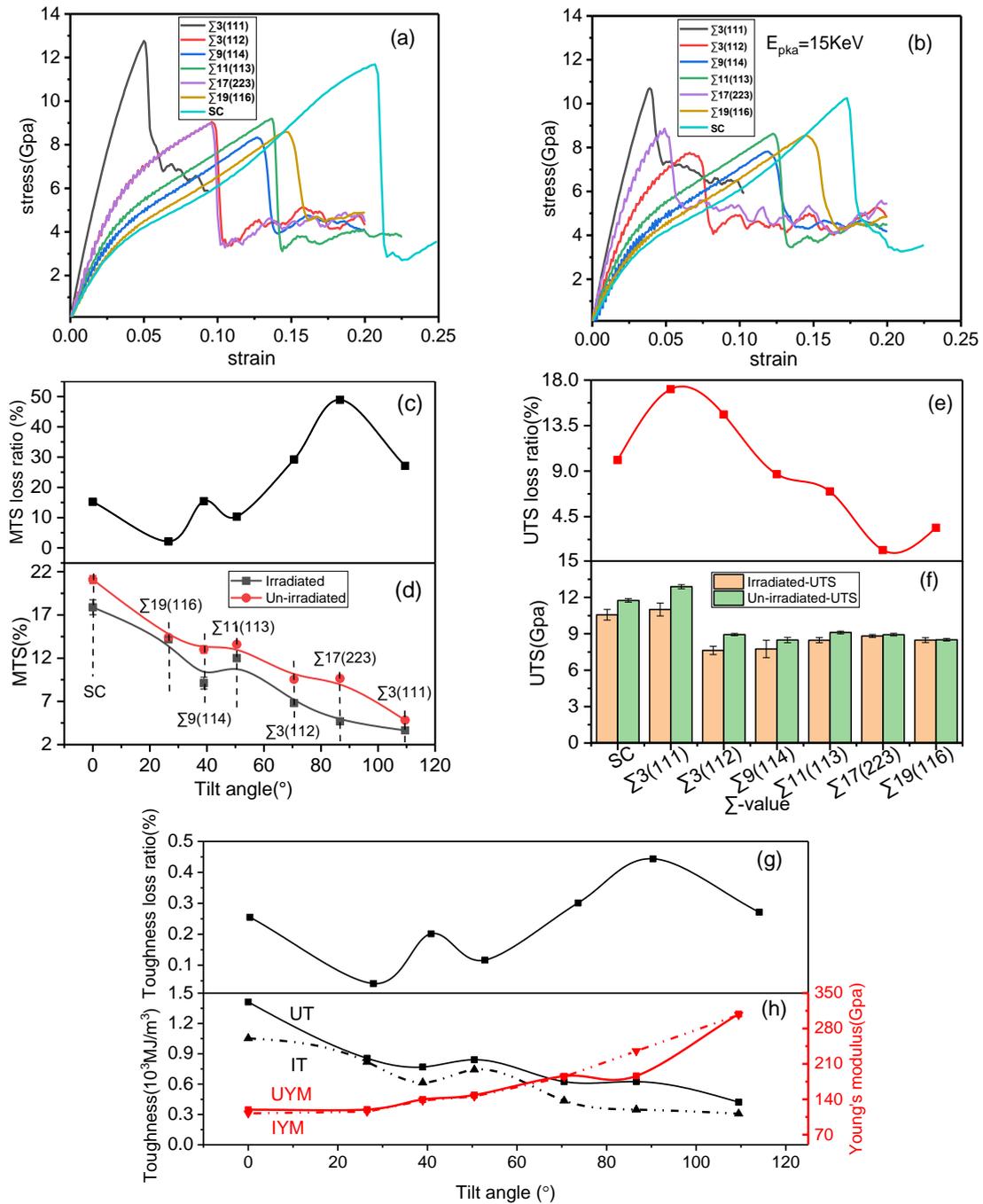

Fig. 7 Tensile stress-strain curves of different GBs (a) before irradiation and (b) after irradiation. (c) MTS loss ratio, and (d) MTS-tilt angle relation, (e) UTS loss ratio, and (f) UTS- Σ values relation, (g) Toughness loss ratio - CSL Σ values relation,(h) Toughness, and Young's modulus of different GBs with and without irradiation. (Note: UT: the toughness of un-irradiated models; IT: the toughness of irradiated models; UYM: Young's modulus of un-irradiated models; IYM: Young's modulus of irradiated models;)

To further explore the effect of irradiation microstructures, the UTS and MTS that relate to tilt angle(θ) and CSL parameter (Σ), are summarized in Fig. 7. The irradiated samples are generally losing ductile and strength compared with the un-irradiated ones (Fig.7d-f). The bi-crystal loses ductility with enhanced tilt angle (Fig. 7d). The UTS-CSL relation is depicted in Fig.7f. The UTS of bi-crystal with small and medium CSL Σ value (*SC, Σ3, Σ9, and Σ11*) lose strength after irradiation; while that with high Σ value (17(223) and Σ19(116)) the strength loss is minor after irradiation. In general, the higher Σ value of GB the unlikely it was influenced by irradiation of the strength. GB itself plays an important role in their mechanical performance difference, irradiated defective microstructures accelerate the failure process. Specifically, Σ19(116) GB has wide GB thickness, excellent ductility, and remarkable strengthen ability before irradiation. Although its thick GB accommodates lots of interstitials, it still holds its ductility and strength after irradiation.

The loss ratio of UTS (MTS) of different GBs after irradiation was calculated by $(R_a - R_b)/ R_a$, where $R_a$ and $R_b$ are the UTS(MTS) of un-irradiated and irradiated GB, respectively. The loss ratio of MTS and UTS of different GBs after irradiation is plotted in Fig. 7(c and e). The order of the loss ratio of UTS follows Σ3(111)> Σ3(112)> Σ9(114)> Σ11(113)> Σ19(116)> Σ17(223), then, from the perspective of strength loss alone, the GBs that have a higher CSL Σ value could maintain a stable strength property when suffering irradiation. In considering the MTS parameter and their loss, Σ9, Σ11, and Σ19 GBs are good in ductility, and could still maintain a stable ductility property when suffering irradiation.

We also measured the toughness of GBs by integrating the tensile stress-strain curve, and the results are shown in Fig. 7h. The irradiated toughness reduce after irradiation, both the un-radiated and radiated toughness is negatively correlated with tilt angle with an exception of Σ11(113), which is quite similar with that of MTS-tilt angle relation(Fig. 7d). The toughness loss ratio is quite similar to the MTS-tilt angle relation(Fig. 7g and Fig. 7c), it is insensitive to irradiation with a medium tilt angle, such as (Σ9 and Σ11). Young's modulus exhibits a positive correlation with the tilt angle, and it seems insensitive to irradiation. Toughness represents the ability of a material to absorb energy without fracturing, then, from the perspective of toughness alone, GB with a medium tilt angle (Σ9 and Σ11), which has considerable toughness and the least loss ratio, is recommended.

Interesting phenomenon has been found for the defect evolution of GBs under tension. Σ3(111) and Σ3(112) GBs with and without irradiation were selected as representatives to illustrate the above phenomenon as shown in Fig. 8. For Σ3(111) without irradiation, the crystal structure is initially dislocation free (Fig. 8(a1)), dislocations nucleate within the grain and evolve during tension (Fig. 8(b1)-(c1)). The Σ3(111) twin boundary hinders the dislocations slip, constraining the dislocations propagating on one side (Fig. 8(c1)). Severe deformation and dense dislocation microstructures cause heavily stress concentration on GB, then, the dislocations begin to break through the GB and slip into another side (Fig. 8(d1)). On the other hand, for Σ3(111) with irradiation, although the residual defects are small, still there some dislocations are created upon irradiation due to defect clusters alongside the GB(Fig. 8(a2)). These dislocations grow on the irradiated side( $1/3 < 111 > \rightarrow 1/6 < 110 > + 1/6 < 112 >$ reaction occurs), cross the GB in a moment, and fill the space with persistent tension(Fig. 8(b2)-(d2)). A large number of stacking fault tetrahedrons are produced upon tension, and the dislocation intersection entanglement is severe, which accounts for the maximal UTS and minimal MTS of Σ3(111) GB sample. Things are alike for Σ3(112), dislocations grow on one side in a small strain and spread over the whole space with tension. Due to the large amount of residual vacancies in GI for the irradiation samples, vacancies aggregate nearby the GB give rise to lots of dislocations (Fig. 8(a4)). These existing dislocations near the GB accelerate the dislocation extension process, thus, leads to ductility loss. In comparison of Σ3(111) and Σ3(112), there is no dislocation entanglement observed in the later, dislocation lines distribute in a specific plane along the specific direction (Fig.8(d3-d4)), and symmetry about GB. The above analysis does provide evidences of the irradiation microstructures' effect on their mechanical performance, the differences depend on the aggregation state of point defects, the distribution and the entanglement degree of dislocations.

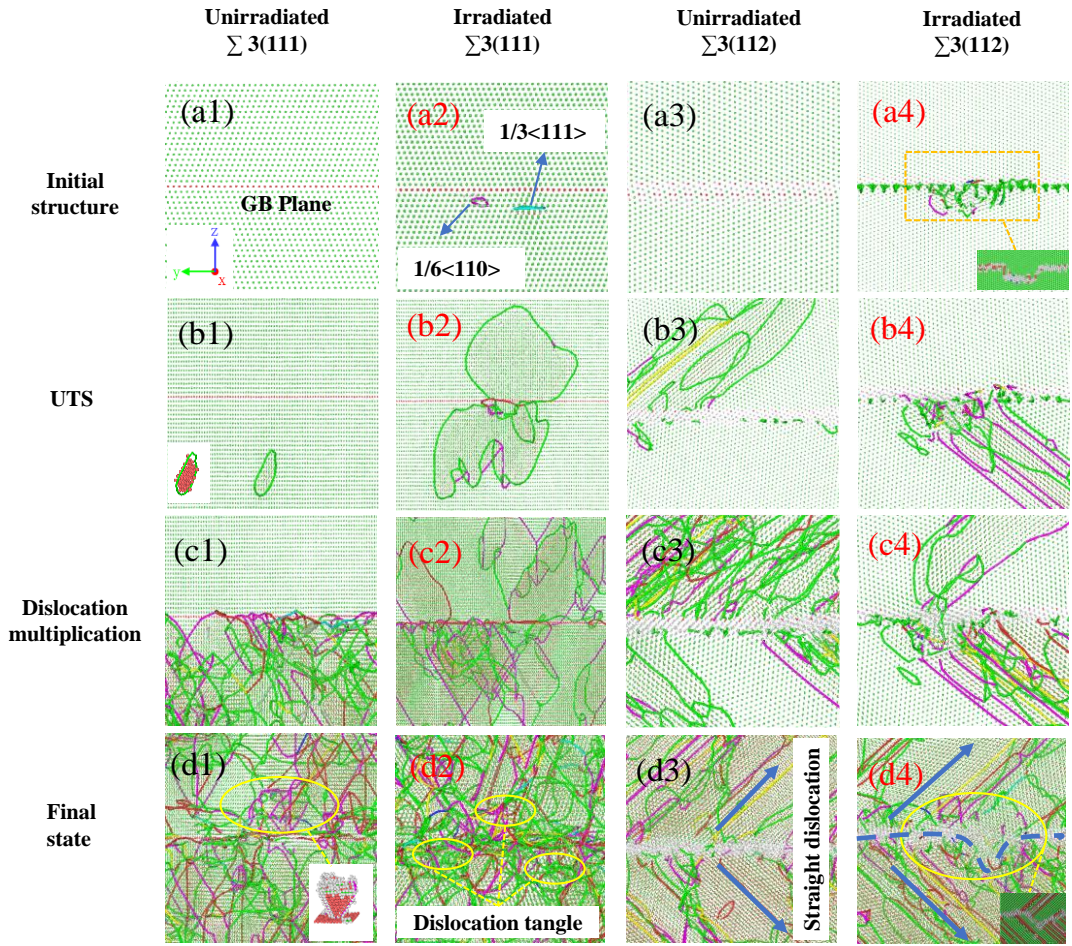

Fig. 8 Dislocation evolution of Σ3(111) and Σ3(112) bi-crystals of their respective un-irradiated and irradiated models during tension. (a1)-(a4) are initial structures before tension, (b1)-(b4) are structures at the corresponding UTS, (c1)-(c4) are dislocation multiplication state under certain plastic deformation stage, and (d1)-(d4) are final state of the tension. Green atoms are of FCC structure, and red for HCP, gray represents structure other than FCC and HCP. Dislocations are distinguished by lines in different colors.

### 3.4 Design strategy for alloys with reinforced irradiation resistance

The rational design of radiation damage-resistant materials should take all kinds of factors into account. In this paper, the irradiation defects and mechanical properties of SC and six kinds of symmetric tilt GB models are compared from four aspects, i.e., toughness, Young's modulus, UTS, and MTS, to select the appropriate radiation damage-resistant materials, as shown in Fig. 9. The comparison of the mechanical properties of the irradiated and un-irradiated crystal shows that SC has high toughness, MTS, and UTS before irradiation. After 15 keV irradiation, although the three properties have decreased, the mechanical properties are still strong compared with other bi-crystal models. Secondly, it should be noted that Σ3(111) bi-crystal has the highest UTS and Young's modulus before and after irradiation, which has strong deformation resistance and high tensile strength limit. It is worth noting that the Σ19(116) and Σ11(113) bi-crystals have high toughness; that is, they have a high potential for deformation energy absorption before fracture.

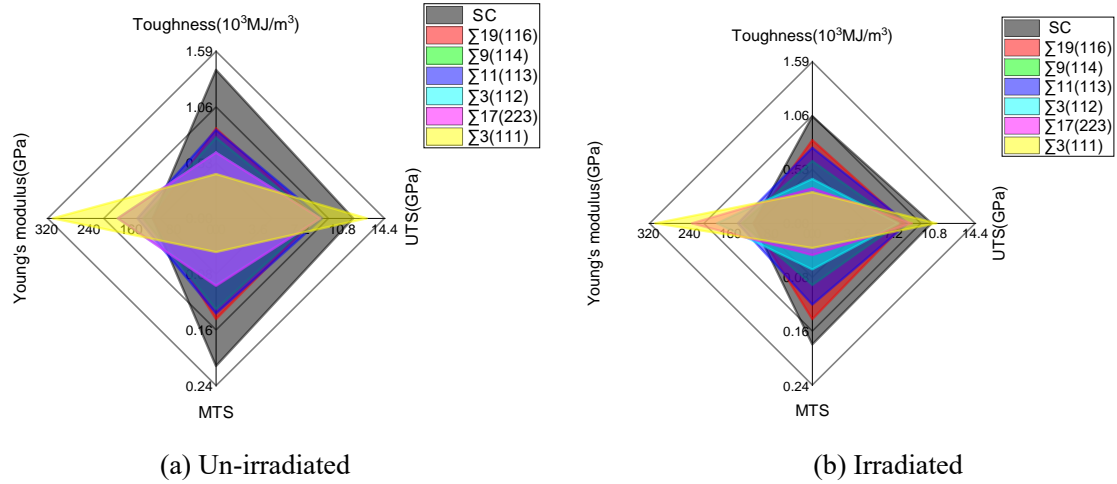

(a) Un-irradiated    (b) Irradiated

Fig. 9 Comparison of mechanical properties of different GB types for (a) with irradiation and (b) without irradiation.

According to the above discussion and comparison, we propose the following design strategies.

(a) Although SC shows remarkable irradiation resistance, the defect combination efficiency should be considered in alloy design, and also the economic efficiency to prepare a single crystal.

(b) To reduce the possibility of voids formation and dislocations propagation caused by the residual vacancies accumulation, GBs with thinner thickness, such as Σ3(111) and Σ11(113), are good choices.

(c) To maintain UTS stability, GBs with medium and high Σ values are better, such as Σ19(116), Σ17(223), Σ11(113), Σ9(114) in turns.

(d) To maintain the ductility stable, GBs with small and medium tilt angle are preferred, such as Σ19(116), Σ9(114), Σ11(113), Σ3(112) in turns.

(e) To get balance toughness and Young's modulus properties, you need to scarify some strength to choose these GBs with the medium tilt angle, such as Σ9(114) and Σ11(113).

## 4 Conclusion

MD calculations were performed to simulate cascade displacement. A respective 2, 5, 10, and 15 keV incident energy were imposed on SC and Σ9(114) bi-crystal to examine the irradiation effect. Tensile stress was applied to SC and Σ3(111), Σ3(112), Σ9(114), Σ11(113), Σ19(116), and Σ17(223) with 15 keV incident energy to investigate the irradiation resistance of various GBs by using stress evolution microstructures and mechanical parameters. The results are drawn as follows.

The irradiation vacancies and interstitials are positively correlated to the incident energy. It is Frankel pairs in SC, and residual vacancies and residual interstitials in the GI of a bi-crystal. Generally, most GBs show obvious bias absorption for interstitials, and a great number of vacancies and limited interstitials are left in GI aggregating into clusters.

Various GBs can be sorted by their GB thickness, tilt angle, CSL parameter $\Sigma$, and the average potential energy. GBs show different defect absorption ability is believed links to these factors. Of all the factors, GB thickness is the most efficient one to measure the irradiation defects, that is, the thicker GB the stronger absorption ability. Strong bias absorption has a dual function, stabilize the microstructures by quick absorption, or accelerate failure once the vacancies are severe aggregated.

Tensile stress microstructures are examined for SC, $\Sigma9(114)$, $\Sigma3(111)$, and $\Sigma3(112)$. Irradiation defect clusters are the origin of dislocations. These dislocations propagate in the whole space without constraint in the SC sample. Whereas in $\Sigma9(114)$, $\Sigma3(111)$, and $\Sigma3(112)$ bi-crystals, dislocations propagation is always restricted to one side of the GB with a smaller strain, GB emits dislocations to the other grain till dislocations wildly propagate.

UTS, MTS, toughness, and Young's modulus are checked after 15keV irradiation. The tilt angle is an important factor for MTS, toughness, and Young's modulus, where the medium tilt angle generally has balanced MTS, toughness, and Young's modulus properties. UTS is measured by the $\Sigma$ value, and a higher one has good UTS stability.

Alloy design should take various GB effects into consideration, and by examining irradiation defects, microstructures, and mechanical performances of different GBs, the present study sheds lights on the design strategies of irradiation-resistant FeNiCr alloy.

**Data Availability**

The raw/processed data required to reproduce these findings cannot be shared at this time due to technical or time limitations.

**CRediT authorship contribution statement**

**Zhijia Liu**: Simulation performance, data analysis and Writing - original draft. **Zehua Feng**: Data analysis, review & editing. **Heran Wang**: Technical support, review & editing. **Yunpeng Zhang**: Review & editing. **Zheng Chen**: Conceptualization, review. **Fanchao Meng**: Data analysis, writing-review & editing. **Jing Zhang**: Conceptualization, writing-review & editing.

**Declaration of competing interest**

The authors declare that there is no conflict of interest associated with this work.

**Acknowledgments**

This Work was supported by the National Natural Science Foundation of China (Grant Nos. 51704243, 52001271, 51674205), and the Natural Science Foundation of Guangdong Province-